\documentclass[
prl%
 ,twocolumn%
 ,secnumarabic%
,amssymb, amsmath,nobibnotes, showpacs,aps, prl]{revtex4}
\usepackage{bm}%
\expandafter\ifx\csname package@font\endcsname\relax\else
 \expandafter\expandafter
 \expandafter\usepackage
 \expandafter\expandafter
 \expandafter{\csname package@font\endcsname}%
\fi
\begin{document}

\title{Leptogenesis from spin-gravity coupling following inflation}
\author{
Subhendra Mohanty$^a$, A.R.Prasanna$^a$, and G. Lambiase$^{b,c}$}
\affiliation{$^a$Physical Research Laboratory, Navrangpura,
Ahmedabad - 380 009, India.}
\affiliation{$^b$Dipartimento di
Fisica "E.R. Caianiello"
 Universit\'a di Salerno, 84081 Baronissi (Sa), Italy.}
 \affiliation{$^c$INFN - Gruppo Collegato di Salerno, Italy.}

\def\nb{\nabla }
\def\gm{\gamma }
\def\al{\alpha }
\def\op{\oplus }
\def\eps{\epsilon }
\def\dg{\dagger }
\def\pr{\prime }
\def\lm {\lambda }
\def\pl{\parallel}
\def\rar{\rightarrow}
\def\be{\begin{equation}}
\def\lan{\left\langle}
\def\ran{\right\rangle}
\def\jo{J_o}
\def\jpm{J_\pm}
\def\j-{\J_-}
\def\i{\item}
\def\ee{\end{equation}}
\def\wt{\tilde}
\def\be{\begin{equation}}
\def\ee{\end{equation}}
\def\al{\alpha}
\def\bea{\begin{eqnarray}}
\def\eea{\end{eqnarray}}
\def\bearr{\begin{eqnarray}}
\def\bearrs{\begin{eqnarray*}}
\def\eearr{\end{eqnarray}}
\def\eearrs{\end{eqnarray*}}
\def\barr{\begin{array}}
\def\earr{\end{array}}
\def\p{\partial}
\def\vb{\vec{b}}
\def\oc{\Omega}
\def\th{\theta}
\def\sg{\sigma}
\def\o{\omega}
\def\bgt{\bigtriangleup}
\def\bgtd{\bigtriangledown}
\def\non\non{\nonumber}
\def\nn8{\nonumber\\[15pt]}
\def\l{\left}
\def\r{\right}
\def\un{\underline}
\def\ve{\varepsilon}
\def\f{\frac}
\def\ts{\textstyle}
\def\dis{\displaystyle}
\def\jv{\vector{J}}
\def\la{\lambda }

\begin{abstract}
The energy levels of the left and the right handed neutrinos is
split in the background of gravitational waves generated during
inflation which, in presence of lepton number violating
interactions, gives rise to a net lepton asymmetry at equilibrium.
Lepton number violation is achieved by the same dimension five
operator which gives rise to neutrino masses after electro-weak
symmetry breaking. A net baryon asymmetry of the same magnitude
can be generated from this lepton asymmetry by electroweak
sphaleron processes.
\end{abstract}
\pacs{04.30.Db, 04.62.+v, 98.80.Cq}
 \maketitle

The successful prediction of light element abundance by the
big-bang nucleosynthesis \cite{bbn} depends on the assumption that
the  net baryon number to entropy ratio $\eta$ as determined by
the recent WMAP \cite{wmap} results is $\eta = 9.2^{+0.6}_{-0.4}
\times 10^{-11}$. Sakharov's well known condition demand that in a
$CPT$ conserving theory, a net lepton/baryon asymmetry can be
generated if there is (a) $B/L$ violating interactions (b) $C$ and
$CP$ violation so that  $B/L$ violating reactions in the forward
and reverse channels do not cancel  and (c) departure from thermal
equilibrium as the statistical distribution of particles and
anti-particles is the same if the Hamiltonian commutes with $CPT$.
If in the theory we have $CPT$ is violation it is possible to
generate lepton/baryon asymmetry at thermal equilibrium  without
requiring $CP$ violation.


In this paper we show that  $CPT$ is violated spontaneously due to
the spin-connection couplings of fermions with cosmological
gravitational waves.  It is well known that inflation
\cite{inflation} generates a nearly scale invariant spectrum of
gravitational waves \cite{grwaves}. The spin connection couplings
split the energy levels of neutrinos compared to anti-neutrinos
and in presence of lepton number violating interaction there is a
net asymmetry generated between neutrinos and anti-neutrinos at
{\it thermodynamic equilibrium}. Lepton number violation is
generated by the dimension five operator introduced by Weinberg
\cite{weinberg} which also generates the neutrino masses after
electro-weak symmetry breaking. The lepton-asymmetry gets
frozen-in when the lepton-number violating processes decouple.
Baryon asymmetry can then be generated from this lepton-asymmetry
by the electro-weak sphaleron processes \cite{ew}. Sphaleron
processes conserve $(B-L)$ so a lepton asymmetry generated in the
GUT era can be converted to baryon asymmetry of the same magnitude
\cite{fukugita}.

The  general covariant coupling of spin $1/2$ particles to gravity
is given by the Lagrangian \cite{kibble} ${\cal{L}} = \sqrt{-g}
(\bar{\psi} \gm^a D_a \psi - m \bar{\psi} \psi )$, $D_a= \p_a -
\f{i}{4} \omega_{bca} \sigma^{bc}$ is the covariant derivative,
and $\omega_{bca}$ are the spin-connections $\omega_{bca} =
 e_{b\lambda}\l(\p_a e^\lambda_{\;\;c} + \Gamma^\lambda_{\gamma
 \mu} e^\gamma_{c} e^\mu_a \r)$.
This Lagrangian  is invariant under the local Lorentz
transformation of the vierbein  $e^a_\mu(x) \rightarrow
\Lambda^a_b(x) e^b_\mu (x)$ and the spinor fields $\psi(x)
\rightarrow \exp(i \epsilon_{a b}(x) \sigma^{a b}) \psi(x)$. Here
$ \sigma^{bc} = \f{i}{2} [\gm^b, \; \gm^c ]$ are generators of
tangent space Lorentz transformation ($a,b,c$ etc. denote the
inertial frame  'flat space' indices and $\al$, $\beta$, $\gamma$
etc. are the coordinate frame 'curved space' indices such that
$e^\mu_a e^{\nu a}= g^{\mu \nu}, e^{a \mu}e^b_\mu= \eta^{ab},
\{\gamma^a, \gamma^b \} = 2 \eta^{a b}$ where $\eta^{a b}$
represents the inertial frame Minkowski metric, and $g_{\mu \nu}$
is the curved space metric).

The spin-connection term in the Dirac equation is a product of three
Dirac matrices
which after some algebra can be reduced to a vector $A^a \gamma_a$
and an axial vector $ i B^d \gamma_5 \gamma_d $. The vector  term
turns out to be anti-hermitian and disappears when the hermitian
conjugate part  is added to the lagrangian ${\cal{L}}$. The
surviving interaction term which describes the spin-connection
coupling of fermions to gravity can be written as a axial-vector
 \bea
 {\cal L}&=& \det(e)~\bar \psi \l(~ i \gamma^a \partial_a~ -~m ~-~  \gamma_5 \gamma_d
 B^d~ \r) \psi\,,
\nonumber\\
 B^d&=&\epsilon^{abcd} e_{b \lambda}  (\partial_a
e^{\lambda}_c + \Gamma^\lambda_{\alpha \mu} e^\alpha_c e^\mu_a
)\,. \label{LI}
 \eea
In a local inertial frame of the fermion, the effect of a
gravitational field appears as a axial-vector interaction
term shown in (\ref{LI}). We now calculate the four vector $B^d$
for a perturbed Robertson-Walker universe.

The general form of perturbations on a flat Robertson-Walker
expanding universe can be written as \cite{bert}
 \bea
 ds^2=a(\tau)^2 [(1+2 \phi) d\tau^2 - \omega_i dx^i d\tau \nonumber\\ -
 ((1+ 2\psi) \delta_{ij} + h_{ij}) dx^i dx^j]\,,
 \label{metric}
 \eea
where $\phi$ and $\psi$ are scalar, $\omega_i$ are vector and
$h_{ij}$ are the tensor fluctuations of the metric. Of the ten
degrees of freedom in the metric perturbations only six are
independent and the remaining four can be set to zero by suitable
gauge choice. For our application we need only the tensor
perturbations and we choose the transverse-traceless (TT) gauge
$h^i _i=0, \partial^i h_{ij}=0$ for the tensor perturbations. In
the TT gauge the perturbed Robertson-Walker metric can be
expressed as
 \bea
 ds^2=a(\tau)^2 [(1+2 \phi) d\tau^2 - \omega_i dx^i d\tau  -
 (1+ 2\psi  - h_{+}) dx_1^2
 \nonumber\\
 -(1+ 2\psi+h_{+})dx_2^2 - 2 h_\times dx_1 dx_2
 -(1+ 2\psi) dx_3^2].
 \label{metric2}
 \eea
\begin{widetext}
An orthogonal set of vierbiens $e^a_\mu$ for this metric is given
by
 \bea
 e_{\mu}^a  = a(\tau) \begin{pmatrix}
  1+ \phi& -\omega_1 & -\omega_2 & -\omega_3 \cr
  0 & -(1+ \psi)+h_+/2  &  h_\times & 0\cr
  0 & 0 & -(1+  \psi)-h_+/2 & 0 \cr
  0& 0& 0& -(1+ \psi)\,\end{pmatrix}.
  \label{vier}
 \eea
\end{widetext}

Using the vierbiens (\ref{vier}) the expression for the components
of the four vector field $B^d$ (\ref{LI}) is given by
 \be\label{B}
 B^0 = \partial_3 h_\times\,, \quad B^i= (\bigtriangledown \times \vec
 \omega)^i+\partial_\tau h_\times\, \delta^{i3}\,.
 \ee
The choice of vierbiens (\ref{vier}) which gives the metric
(\ref{metric}) is not unique as one can make a local Lorentz
transformation (LLT) ${e^m_\mu }^\prime = {\Lambda^m_n}^\prime
~e^n_\mu$. Under a LLT the four-vector $B^d$ transforms as
${B^d}^\prime = {\Lambda^d_a}^\prime~B^a$. The dispersion
relations of the left and the right helicity fermions have $B^d$
dependent terms of the form $\eta_{m n} ~B^m B^n$ and $\eta_{m
n}p^m B^n$, and therefore the dispersion relations do not change
with a transformation of the local inertial frame.

The fermion bilinear term $\bar \psi \gamma_5 \gamma_a \psi $ is
odd under $CPT$ transformation. When one treats $B^a$ as a
background field  then the interaction term in (\ref{LI})
explicitly violates $CPT$. When the primordial metric fluctuations
become classical, i.e  there is no back-reaction of the
micro-physics involving the fermions on the metric and $B^a$ is
considered as a fixed external field, then CPT is violated
spontaneously.

The gravitational spin connection coupling for the neutrinos at
high energy is given by
 \bea \label{Lnu}
  {\cal L}&=&\det(e) [\left(i{\bar \nu_L}\gamma^a \partial_a\nu_L+i{\bar
  \nu_R}\gamma_a \partial_a \nu_R \right) \\
   &+& m {\bar \nu_L} \nu_R  +m^\dagger  {\bar \nu_R} \nu_L
   +  B^a (\bar \nu_R \gamma_a \nu_{R}-\bar
   \nu_{L}\gamma_a\nu_L)]\,, \nonumber
 \eea
where $B^a$ are the parameters of the gravitational waves as
defined in $(\ref B)$. If we consider only the Standard Model
fermions then the right handed neutrinos carry the opposite Lepton
number compared to the left handed neutrinos, $\nu_R = (\nu_L)^c$
and the mass term in (\ref{Lnu}) is of the Majorana type (we have
suppressed the generation index).

The dispersion relation of left and right helicity neutrinos
fields are given by $\eta^{a b}(p_a + \xi B_a)(p_b + \xi
B_b)=m^2$, where $\xi=-1$ for $\nu_L$ and $\xi=1$ for $\nu_R$.
Keeping terms linear in the perturbations $B^a$, the free particle
energy of the left and right  helicity states is
 \be
 E_{L,R}(p) =p + \frac{m^2}{2 p}  \mp \left(B_0 -\frac{{\bf p} \cdot {\bf
 B}}{p}\right)\,,
 \label{energy}
 \ee
with $p=|{\bf p}|$. In the Standard Model $\nu_L$ carry lepton
number $+1$ and $\nu_R$ are assigned lepton number $(-1)$. In the
presence of non-zero metric fluctuations, there is a split in
energy levels of $\nu_{L,R}$ given by (\ref{energy}). If there are
GUT processes that violate lepton number freely above some
decoupling temperature $T_d$, then the equilibrium value of lepton
asymmetry generated for all $T>T_d$ will be
 \be
 n(\nu_L)-n(\nu_R)= \frac{g}{2 \pi^2} \int d^3p
 \left[\frac{1}{1+
 e^{\frac{E_{L}}{T}}}-\frac{1}{1+e^{\frac{E_{R}}{
T}}}\right] \label{therm1}
 \ee
The spin-connection coupling with gravitational waves also splits
the energy levels between the charged left and right handed
fermions. For example $E(e^-_R) -E(e^-_L)=2 (B_0 -{\bf p} \cdot
{\bf B}/p)$. But this does not lead to lepton generation of lepton
asymmetry as both $e^-_L$ and $e^-_R$ carry the same lepton
number.

In the ultra-relativistic regime $p \gg m_\nu $ and assuming that
$B_0 \ll T$, the expression (\ref{therm1}) for lepton asymmetry
reduces to
 \be
 \Delta n_L = \frac{g T^3}{6} \l(\frac{B_0}{T}\r)\,.
 \label{deltan}
 \ee
The dependence on ${\bf B}$ drops out after angular integration in
(\ref{therm1}) and the lepton asymmetry depends  on the tensor
perturbations only through $B^0$.

To compute the spectrum of gravitational waves $h({\bf x},\tau)$
during inflation, we  express $h_{\times}$ in terms of the
creation- annihilation operator
 \be
 h({\bf x}, \tau)=\frac{\sqrt{16 \pi}}{a M_p}\int \frac{d^3 k}
 {(2\pi)^{3/2}}
 \left(a_{\bf k} ~f_k(\tau) +a^{\dagger}_{-{\bf k}}~f_{k}^*(\tau)
 \right)e^{i{\bf  k} \cdot {\bf x}}
 \label{h}
 \ee
where ${\bf k}$ is the comoving wavenumber, $k=|{\bf k}|$, and
$M_p= 1.22 \,\,10^{19}GeV$ is the Planck mass. The mode functions
$f_k(\tau)$ obey the minimally coupled Klein-Gordon equation
 \be
 f_k^{\prime \prime} + \left(k^2 -
 \frac{a^{\prime \prime}}{a}\right)f_k=0\,. \label{f2}
 \ee
During de Sitter era, the scale factor $a(\tau)=-1/(H_I \tau)$
where $H_I$ is the Hubble parameter, and Eq.~(\ref{f2}) has the
solution
 \be
 f_k( \tau)= \frac{e^{-i k \tau}}{\sqrt{2 k}}\left(1- \frac{i}{k
  \tau}\right)\,, \label{sol1}
 \ee
which matches the positive frequency "flat space" solutions $e^{-i
k \tau}/\sqrt{2 k}$ in the limit of $k \tau \gg 1$. The first term
of (\ref{sol1}) represents the decaying part of $h$ and can be
dropped. The second term of (\ref{sol1}) represents the amplitude
constant gravitational wave, which survives to the present era.
Substituting the second term of (\ref{sol1}) in in (\ref{h}) and
using the canonical commutation relation for $a_k$ and
$a_k^\dagger$ we get the standard expression for two point
correlation of gravitational waves generated by inflation \be
 \langle h({\bf x},\tau)h({\bf x},\tau)\rangle^{inf} \equiv \int \frac{dk}
{k}\, (|h_k|^2)^{inf}\,,
 \ee
with the spectrum  of gravitational waves
 given by the scale invariant form
 \be
 (|h_k|^2)^{inf}=\frac{4}{\pi} \frac{H_I^2}{M_p^2} \,.
\label{hinf}
 \ee
There is a stringent constraint $H_I/M_p < 10^{-5}$ from CMB data
\cite{krauss}. This constraint limits the parameter space of
interactions that can be used for generating the requisite
lepton-asymmetry. In the radiation era, when these modes re-enter
the horizon, the amplitude redshifts by $a^{-1}$ from the time of
re-entry. The reason is that in the radiation era $a(\tau) \sim
\tau$ and the equation for $f_k$, (\ref{f2}), gives plane wave
solutions $f_k= (1/\sqrt{2k})exp(-i k \tau)$. Therefore in the
radiation era the amplitudes of $h$ redshifts as $a^{-1}$. The
gravitational waves inside the horizon in the radiation era will
be
 \be
h^{rad}_k= h^{inf}_k \,\frac{a_k}{a(\tau)}=h^{inf}_k
\,\frac{T}{T_k}\,, \label{hrad}
 \ee
where $h^{inf}_k$ are the gravitational waves generated by
inflation (\ref{hinf}), $a_k$ and $T_k$ are the scale factor and
the temperature when the modes of wavenumber $k$ entered the
horizon in the radiation era. The horizon entry of mode $k$ occurs
when
 \be
 \frac{a_k H_k}{k}= \frac{a(T)\,T\, H_k}{T_k\,k}=1\,,
 \label{hc}
 \ee
where $H_k= 1.67 \sqrt{g_*} T_k^2 /M_p$ is the Hubble parameter at
the time of horizon crossing of the $k$ the mode ($g_*$ is the
number of relativistic degrees of freedom which for the Standard
Model is $g_*=106.7$). Solving equation (\ref{hc}) for $T_k$ we
get \vskip -0.8 truecm
 \be
 T_k= \frac{1}{1.67 \sqrt{g_*}}\frac{k\,M_p}{a(\tau)\,T}
 \,.\label{Tk}
 \ee
The amplitude of the gravitational waves of mode $k$ inside the
radiation horizon is, using (\ref{Tk}) and (\ref{hrad}), given by
$h^{rad}_k= h^{inf}_k \,\frac{a(T)}{k}\,\frac{T^2\, 1.67\,
\sqrt{g_*}}{M_p}$.
Note that the gravitational wave spectrum inside the radiation era
horizon is no longer scale invariant. The gravitational waves in
position space have the correlation function
 \be
 \langle h({\bf x},\tau)h({\bf x},\tau)\rangle^{rad} =\int \frac{dk}
{k}\, (h^{rad}_k)^2\,,
 \ee
and hence for the spin connection $B^0$ generated by the
inflationary gravitational waves in the radiation era, we get
 \bea\label{B2}
 \langle B^0({\bf x},\tau)B^0({\bf x},\tau)\rangle &=&\int
\frac{dk}{k}\,\left(\frac{k}{a}\right)^2\,(h^{rad}_k)^2 \\
 &=&\frac{4}{\pi} \left(\frac{H_I}{M_p^2} \, \, T^2\, 1.67\,
\sqrt{g_*}\right)^2 \int_{k_{min}}^{k_{max}} \frac{dk}{k} \nonumber\\
 \nonumber
 \eea
 \vskip -0.4 truecm
Now we find that the spectrum of spin-connection is scale
invariant inside the radiation horizon. This is significant in
that  the lepton asymmetry generated by this mechanism depends
upon the infrared and ultraviolet scales only logarithmically. The
scales outside the horizon are blue-tilted which means that there
will be a scale dependent anisotropy in the lepton number
correlation at two different space-time points $\langle \Delta
L(r)\Delta L(r^\prime) \rangle \sim A \,k^n\, , n>0$, where
$\Delta L(r) \equiv L(r)-\bar L$, is the anisotropic deviation
from the mean value. Unlike in the case of CMB, this anisotropy in
the lepton number is unlikely to be accessible to experiments.
Nucleosynthesis calculations only give us an average value at the
time of nucleosynthesis (when $T \sim 1 MeV$). The maximum value
of $k$ are for those modes which leave the de Sitter horizon at
the end of inflation. If inflation is followed by radiation
domination era starting with the re-heat temperature $T_{RH}$ then
the maximum value of $k$ in the radiation era (at temperature $T$)
is given by $ k_{max}/a(T)= H_I \left(\frac{T}{T_{RH}}\right)$.
The lower limit of $k$ is $k_{min}=e^{-N}k_{max}$ which are the
modes which left the de-Sitter horizon in the beginning of
inflation ($N$ is the total e-folding of the scale factor during
inflation, $N\simeq 55-70$). The integration over $k$ then yields
just the factor $\ln(k_{max}/k_{min})=N$. The r.m.s value of spin
connection that determines the lepton asymmetry through equation
(\ref{deltan}) is $(B_0)_{rms}=\sqrt{ \langle B_0^2\rangle}$,
 \be
(B_0)_{rms} = \frac{2 }{\sqrt{\pi}}\left(\frac{H_I}{M_p^2} \, \,
T^2\, 1.67\, \sqrt{g_*}\right)\,\sqrt{N}\,.
 \ee
The lepton asymmetry (\ref{deltan})  as a function of temperature
can therefore be expressed as (taking $g= 3$ for the three
neutrino flavors)
 \be
 \Delta n_L(T) =\frac{1}{\sqrt{\pi}}\,(1.67\, \sqrt{g_*})\, \sqrt{N}
 \left(\frac{T^4\, H_I} {M_p^2}\right)\,.
 \ee
The lepton number to entropy density ($s=0.44~ g_*~ T^3$ )  is
given by
 \be
 \Delta L \equiv \frac{\Delta n_L(T)}{s(T)}
 \simeq 2.14 \,  \frac{T\,H_I \,\sqrt{N}}{ M_{p}^2\,\sqrt{g_*}}\,.
\label{L}
 \ee
Lepton number asymmetry will be generated as long as the lepton
number violating interactions are in thermal equilibrium. Once
these reactions decouple at some decoupling temperature $T_d$,
which we shall determine, the $\Delta n_L(T)/s(T)$ ratio remains
fixed for all $T<T_d$.

To calculate the decoupling temperature of the lepton number
violating processes we turn to a specific effective dimension five
operator which gives rise to Majorana masses for the neutrinos
introduced by Weinberg \cite{weinberg} ${\cal{L}}_W=
\frac{C_{\alpha \beta} }{2M}(\overline{ {l_{L
 \alpha}}^c }~ \tilde{\phi^*})(\tilde{\phi^\dagger}~ l_{L \beta}) +
 h.c.$
where $l_{L \alpha}= (\nu_\alpha , e^-_\alpha)_L^T $ is the
left-handed lepton doublet ($\alpha $ denotes the generation),
$\phi=(\phi^+,\phi^0)^T $ is the Higgs doublet and  $\tilde
\phi\equiv i \sigma_2 \phi^*=(-{\phi^0}^*, \phi^-)^T$.
$M$ is some large mass scale and $C_{\alpha \beta}$ are of order
unity.

The $\Delta L=2$ interactions that result from the operator
${\cal{L}}_W$ are
 \be
 \nu_L + \phi^0 \leftrightarrow \nu_R + \phi^0\,, \quad
 \nu_R + {\phi^0}^*
 \leftrightarrow \nu_L + {\phi^0}^*\,.
 \label{ints}
 \ee
In the absence of the gravitational waves the forward reactions
would equal the backward reactions and no net lepton number would
be generated. In the presence of a background gravitational waves
the energy levels of the left and right helicity neutrinos are no
longer degenerate and this leads to a difference in the number
density of left and right handed neutrinos of the magnitude given
by equation (\ref{deltan}) at thermal equilibrium. This process
continues till the interactions (\ref{ints}) decouple. The
decoupling temperature is estimated as follows. The cross section
for the interaction $\nu_{L \alpha} +\phi^0 \leftrightarrow \nu_{R
\beta} +\phi^0$ is $ \sigma=\frac{|C_{\alpha \beta}|^2}{M^2}
\frac{1}{\pi}$, and interaction rate $\Gamma =\langle n_\phi~
\sigma\rangle $ of the $\Delta L=2$ interactions is $\Gamma =
\frac{0.122}{\pi} \frac{|C_{\alpha\beta}|^2 T^3}{M^2}$. In the
electroweak era, when the Higgs field in ${\cal{L}}_W$ acquires a
$vev$, $\langle\phi\rangle =(0,v)^T$ (where $v=174~ GeV$), this
operator gives rise to a neutrino mass matrix $m_{\alpha \beta}=
\frac{v^2 C_{\alpha \beta}}{M}$. We can therefore substitute the
couplings $\frac{C_{\alpha \beta}}{M}$ in terms of the light left
handed Majorana neutrino mass. At the decoupling temperature the
interaction rate $\Gamma (T)$ falls below the expansion rate
$H(T)= 1.7 \sqrt{g_*} ~T^2/M_{p}$. The decoupling temperature is
obtained from equation $\Gamma (T_d)=H(T_d)$
and turns out to be \vskip -0.9 truecm
 \be
 T_d= 13.68 \pi~ \sqrt{g_*}~ \frac{v^4}{m_\nu^2 M_{p}}\,,
\label{Td}
 \ee
where $m_\nu$ is the mass of the heaviest neutrino. A lower bound
on the  mass of the heaviest neutrino is given by atmospheric
neutrino experiments \cite{super-K} $m_\nu^2 > \Delta_{atm}= 2.5
\,\, 10^{-3} eV^2$ which means that the decoupling temperature has
an upper bound given by $ T_d = 1.3\,\, 10^{13}
(\Delta_{atm}/m_\nu^2) \, GeV\,$. Substituting the expression
(\ref{Td}) for $T$ in (\ref{L}), we finally obtain the formula for
lepton number \vskip -0.9 truecm
 \bea
 L&=& 92.0 \,\left(\frac{v^4 \, H_I}{m_\nu^2 \, M_p^3}\,\right)\,\sqrt{N} \nonumber\\
 &=& 7.4 \, 10^{-11} \frac{H_I}{4 \, 10^{14} GeV}\,\frac{2.5 \, 10^{-3}
 eV^2}{m_\nu^2}\,\frac{\sqrt{N}}{10}\,.
 \eea
 As first pointed out in \cite{fukugita}, electroweak spahalerons at the
temperatures $T \sim 10^3 GeV$  violate $B+L$ maximally and
conserve $B-L$. Therefore a lepton asymmetry generated at an
earlier epoch gets converted to baryon asymmetry of the same
magnitude by the electroweak sphalerons.

The input parameters needed for generating the correct magnitude
of baryogenesis ($\eta \sim 10^{-10}$) are the amplitude of the
$h_\times\sim 10^{-6}$ (or equivalently the curvature during
inflation $H_I \sim 10^{14} GeV$ or the scale of inflation is the
GUT scale, $V^{1/4} \sim 10^{16} GeV$ which is allowed by CMB
\cite{pilo,Knox}, neutrino Majorana mass in the atmospheric
neutrino scale $m_{\nu}^2 \sim 10^{-3} eV^2$ \cite{super-K} and
duration of inflation $H_I t =N \sim 100$ needed to solve the
horizon and entropy problems in the standard inflation paradigm.
All these parameters are well within experimentally acceptable
limits.


To summarize the mechanism of baryogenesis we propose arises in
the standard Einstein's gravity where spontaneous  $CPT$ violation
is caused by gravitational waves,
and the $h_\times$ gravitational wave modes which give non-zero
spin connection are produced in {\it generic} inflation scenarios
(in contrast to models \cite{alexander} where  baryon asymmetry is
created through a gravitational Chern-Simons term, which can be
generated if specific  $CP$ violating terms are introduced in the
inflaton potential which can give rise to birefringent circularly
polarized gravitational waves).


\end{document}